\documentclass[pra,aps,twocolumn,showpacs]{revtex4} 
\usepackage{graphicx}
\newcommand{\ket}[1]{|#1\rangle} 
\newcommand{\bra}[1]{\langle #1|} 
\newcommand{\Tr}{\text{Tr}} 
\begin{document} 
\title{Kinematic approach to off-diagonal geometric phases of 
nondegenerate and degenerate mixed states} 
\author{D.M. Tong$^1$, Erik Sj\"{o}qvist$^2$\footnote{Electronic       
address: eriks@kvac.uu.se}, Stefan Filipp$^{3,4}$\footnote{Electronic       
address: sfilipp@ati.ac.at}, L.C. Kwek$^{1,5}$, 
and C.H. Oh$^1$\footnote{Electronic address: phyohch@nus.edu.sg}}       
\affiliation{$^1$Department of Physics, National University of       
Singapore, 10 Kent Ridge Crescent, Singapore 119260, Singapore \\       
$^2$Department of Quantum Chemistry, Uppsala University, Box 518,   
Se-751 20 Uppsala, Sweden \\  
$^3$Atominstitut der \"Osterreichischen Universit\"aten, Stadionallee 2,  
A-1020 Vienna, Austria\\  
$^4$Institut Laue Langevin, Bo\^ite Postale 156, F-38042 Grenoble
Cedex 9, France\\     
$^5$National Institute of Education, Nanyang Technological       
University, 1 Nanyang Walk, Singapore 639798, Singapore}  
\date{\today}       
\begin{abstract}   
Off-diagonal geometric phases have been developed in order to  
provide information of the geometry of paths that connect  
noninterfering quantal states. We propose a kinematic approach  
to off-diagonal geometric phases for pure and mixed states. We  
further extend the mixed state concept proposed in [Phys. Rev.  
Lett. {\bf 90}, 050403 (2003)] to degenerate density operators.  
The first and second order off-diagonal geometric phases are 
analyzed for unitarily evolving pairs of pseudopure states. 
\end{abstract} 
\maketitle       
\date{\today}       
\section{introduction}  
Pancharatnam's \cite{Pancharatnam} work on the geometric phase in  
interference of light waves in distinct states of polarization   
plays a prominent role among early anticipations of the concept  
of geometric phase. It found its quantal counterpart when Berry  
\cite{Berry} discovered geometric phase factors accompanying  
cyclic adiabatic changes. Since then there has been an immense 
interest in holonomy effects in quantum mechanics, which has led to 
many generalizations of the notion of geometric phase. The extension 
to nonadiabatic cyclic evolution was developed by Aharonov and Anandan 
\cite{aharonov87,anandan88}. Samuel and Bhandari \cite{Samuel} generalized 
the pure state geometric phase further by extending it to noncyclic 
evolution and sequential projection measurements.  Mukunda and Simon 
\cite{Mukunda} put forward a kinematic approach to the geometric 
phase. Further generalizations and refinements, by relaxing the 
conditions of adiabaticity, unitarity, and the cyclic nature of the 
evolution, have since been carried out \cite{pati95a,pati95b}. 
  
The concept of geometric phases for pure states has been extended 
to mixed states. This was first addressed by Uhlmann 
\cite{uhlmann86,uhlmann91} within the mathematical context of
purification.  Sj\"oqvist {\it et al.} \cite{sjoqvist00} introduced an
alternative definition of geometric phase for density operators with
all the nonzero eigenvalues being nondegenerate, based upon the
experimental context of quantum interferometry. Singh {\it et al.} 
\cite{singh03} gave a kinematic description of the mixed state
geometric phase in Ref.
\cite{sjoqvist00} and extended it to degenerate density operators. 
The generalization to nonunitary evolution has also been addressed 
in Refs. \cite{ericsson03,peixoto03,tong04} and experimental test 
of the phase concept in Ref. \cite{sjoqvist00} has been carried out  
using nuclear magnetic resonance \cite{du03}.  
  
All the above notions of geometric phase break down in cases where the
interference visibility between the initial and final states vanishes
leading to an interesting nodal point structure in the experimental
parameter space that could be monitored in a history-dependent manner
\cite{bhandari91,bhandari92a,bhandari92b,bhandari93,bhandari00,anandan00}. 
This problem may be overcome by introducing the 
concept of off-diagonal geometric phases, as was first put forward 
for pure states in adiabatic evolution by Manini and Pistolesi
\cite{manini00}. Mukunda {\it et al.} \cite{Mukunda2} removed the  
adiabaticity assumption of Ref. \cite{manini00} and Hasegawa {\it et
al.} \cite{hasegawa01,hasegawa02} verified the off-diagonal pure state
geometric phase by a neutron experiment. The concept of off-diagonal
geometric phase has been extended to mixed states
\cite{filipp03a,filipp03b} represented by nondegenerate density
operators, as well as to the Uhlmann holonomy \cite{filipp04}. These
off-diagonal mixed state geometric phases contain information about
the geometry of state space along the path connecting pairs of density
operators, when the standard mixed state geometric phases in Refs.
\cite{uhlmann86,uhlmann91,sjoqvist00} are undefined. In this
paper, we propose a kinematic approach to the off-diagonal mixed state
geometric phase in Refs. \cite{filipp03a,filipp03b} and further extend 
it to the degenerate case.
  
\section{kinematic approach to off-diagonal geometric phases}  
Consider an $N$ dimensional quantum system $s$, the pure states of
which represented by vectors of Hilbert space ${\cal H}_s$. Let
$\ket{\psi_1},\ldots,\ket{\psi_N}$ denote an arbitrary orthonormal
basis of ${\cal H}_s$ and $\{ U^{\parallel}(t) |
t\in[0,\tau],U^{\parallel}(0)=I \}$ be a one-parameter family of
unitarities fulfilling the parallel transport conditions
\cite{sjoqvist00} 
\begin{eqnarray}       
\bra{\psi_k} U^{\parallel\dagger} (t)       
\dot{U}^{\parallel} (t) \ket{\psi_k} = 0 , \ k=1,\ldots,N   
\label{parallel}       
\end{eqnarray}       
relative to the chosen basis. For any pair of basis vectors  
$\ket{\psi_{j_1}}$ and $\ket{\psi_{j_2}}$ such that 
$\bra{\psi_{j_1}} U^{\parallel} (\tau) \ket{\psi_{j_2}} \neq 0$  
and $\bra{\psi_{j_2}} U^{\parallel} (\tau) \ket{\psi_{j_1}} \neq 0$, 
the quantity  
\begin{eqnarray}  
\gamma_{j_1 j_2}^{(2)} =  
\Phi \Big[ \bra{\psi_{j_1}} U^{\parallel} (\tau)  
\ket{\psi_{j_2}} \bra{\psi_{j_2}} U^{\parallel} (\tau)  
\ket{\psi_{j_1}} \Big] ,    
\label{gammap}  
\end{eqnarray}  
where $\Phi[z] \equiv z/|z|$ for any nonzero complex number $z$, 
is gauge invariant and therefore a property of the pair of paths  
${\cal C}_{j_1}, {\cal C}_{j_2}: t\in [0,\tau] \rightarrow  
\ket{\psi_{j_1}(t)} \bra{\psi_{j_1}(t)},\ket{\psi_{j_2}(t)}  
\bra{\psi_{j_2}(t)}$ in projective Hilbert space. The set  
$\{ \gamma_{j_1 j_2}^{(2)} \}$ constitutes the second order  
off-diagonal pure state geometric phase factors \cite{manini00}.  
It may be extended by considering $l\leq N$ states yielding  
the $l$th order off-diagonal pure state geometric phase factors  
\begin{eqnarray}  
\gamma_{j_1\ldots j_l}^{(l)} =  
\prod_{a=1}^l \Phi \left[ \bra{\psi_{j_a}} U^{\parallel} (\tau)  
\ket{\psi_{j_{a+1}}} \right] ,  
\label{gamma3}  
\end{eqnarray}  
where $\ket{\psi_{j_{l+1}}}=\ket{\psi_{j_1}}$ and 
$\bra{\psi_{j_a}} U^{\parallel} (\tau)  
\ket{\psi_{j_{a+1}}} \neq 0$ for all $a=1,\ldots l$.  
  
Notice that there is an equivalence set ${\cal S}$ of unitarities  
$\widetilde{U}(t)$ that all realize $\{ {\cal C}_k \}$, namely  
those of the form  
\begin{eqnarray}       
\widetilde{U}(t) =       
U(t) \sum\limits_{k=1}^N e^{i\theta_k (t)} 
\ket{\psi_k} \bra{\psi_k} ,       
\label{U}       
\end{eqnarray}       
where $U(t) \in {\cal S}$ and $\theta_k (t)$ are real time-dependent 
parameters such that $\theta_k (0)=0$.  We may in particular identify 
$U^{\parallel} (t) \in {\cal S}$ by substituting $U^{\parallel} (t) = 
\widetilde{U} (t)$ into Eq. (\ref{parallel}), so that we obtain   
\begin{eqnarray}      
\theta_k (t) \equiv \theta_k^{\parallel} (t) = 
i \int_0^t \bra{\psi_k} U^{\dagger} (t') \dot U (t') \ket{\psi_k} dt'        
\end{eqnarray}  
and   
\begin{eqnarray}       
U^\parallel(t) =       
U(t) \sum\limits_{k=1}^N e^{-\int_0^t \bra{\psi_k} U^{\dagger} (t')      
\dot U (t') \ket{\psi_k} dt'}    
\ket{\psi_k} \bra{\psi_k} .        
\label{Uparallel}       
\end{eqnarray}  
Inserting this expression into Eq. (\ref{gammap}) for 
$U^{\parallel} (t)$, we obtain a kinematic expression 
for the $l=2$ off-diagonal geometric phase factors as  
\begin{eqnarray}       
\gamma_{j_1 j_2}^{(2)} & = &  
\Phi \Big[ \bra{\psi_{j_1}} U(\tau) \ket{\psi_{j_2}} \bra{\psi_{j_2}}  
U(\tau) \ket{\psi_{j_1}} \Big] 
\nonumber \\ 
 & & \times e^{-\int_0^\tau \bra{\psi_{j_1}}  
U^{\dagger} (t) \dot{U} (t) \ket{\psi_{j_1}} dt}  
\nonumber \\ 
 & & \times e^{-\int_0^\tau  
\bra{\psi_{j_2}} U^{\dagger} (t) \dot{U} (t) \ket{\psi_{j_2}} dt}.  
\label{gamma2}       
\end{eqnarray}        
In the above expression, the unitary operator $U(t) \in \mathcal{S}$ 
need not satisfy the parallel transport conditions. One may verify  
that $\gamma_{j_1 j_2}^{(2)}|_{\widetilde{U} (t)} =  
\gamma_{j_1 j_2}^{(2)}|_{U(t)}$ for any choice of $\theta_k (t)$. 
  
An interesting aspect of Eq. (\ref{gamma2}) is that in the qubit  
case with $\ket{\psi_1} = \ket{0}$ and $\ket{\psi_2} = \ket{1}$,  
it follows that $\bra{0} U^{\dagger} (t) \dot{U} (t) \ket{0} = -  
\bra{1} U^{\dagger} (t) \dot{U} (t) \ket{1}$ for any SU(2) operation,  
leading to  
\begin{eqnarray}  
\gamma_{01}^{(2)} & = & \Phi \Big[ \bra{0} U(\tau) \ket{1} \bra{1}  
U(\tau) \ket{0} \Big] 
\nonumber \\ 
 & = & \Phi \Big[ \bra{0} U^{\parallel}(\tau) \ket{1}  
\bra{1} U^{\parallel}(\tau) \ket{0} \Big] = -1,   
\end{eqnarray}  
whenever $\bra{0} U(\tau) \ket{1} \neq 0$ and 
$\bra{1} U(\tau) \ket{0} \neq 0 $. Thus, the $\pi$ shift in the 
qubit case is independent of the fulfillment of the parallel 
transport condition Eq. (\ref{parallel}); a result that has been 
experimentally verified for neutron spin \cite{hasegawa01,hasegawa02}.  
  
Similarly, for the higher order off-diagonal geometric phases  
\cite{manini00}, we obtain      
\begin{eqnarray}       
\gamma_{j_1 \ldots j_l}^{(l)} & = & \prod_{a=1}^{l}  
\Phi \Big[ \bra{\psi_{j_a}} U(\tau) \ket{\psi_{j_{a+1}}}  
\Big] 
\nonumber \\ 
 & & \times e^{-\int_0^\tau \bra{\psi_{j_a}}  
U^{\dagger} (t) \dot{U} (t) \ket{\psi_{j_a}} dt} , \ l \leq N 
\label{gamma4}       
\end{eqnarray}   
upon substitution of Eq. (\ref{Uparallel}) into Eq. (\ref{gamma3}).  
Once again, one may verify that 
$\gamma_{j_1\ldots j_l}^{(l)}|_{\widetilde{U} (t)} =  
\gamma_{j_1\ldots j_l}^{(l)}|_{U(t)}$.
  
We now put forward a kinematic approach to the off-diagonal 
geometric phases of nondegenerate mixed states proposed in 
Refs. \cite{filipp03a,filipp03b}.  
Let  
\begin{eqnarray}  
\rho_1 = \lambda_1 \ket{\psi_1}\bra{\psi_1} + \ldots + 
\lambda_N \ket{\psi_N} \bra{\psi_N} , 
\label{rho1} 
\end{eqnarray}  
where for all nonzero eigenvalues we have $\lambda_k \neq 
\lambda_{l\neq k}$. Furthermore, introduce the unitarity  
\begin{eqnarray}  
W = \ket{\psi_1}\bra{\psi_N} + \ket{\psi_N}\bra{\psi_{N-1}} + 
\ldots + \ket{\psi_2}\bra{\psi_1}   
\label{eq:W}
\end{eqnarray}  
in terms of which any pair of members $\rho_{j_1}$, $\rho_{j_2}$, 
$j_1 \neq j_2$, of the set  
\begin{eqnarray} 
\rho_n = W^{n-1} \rho_1 \big( W^{\dagger} \big)^{n-1},  n=1,\ldots,N
\label{eq:rhon} 
\end{eqnarray}  
are unitarily connected and do not interfere since $\Tr \big( W^{j_2 -
j_1} \rho_{j_1} \big)$ vanishes. In analogy with the pure state case,
the density operators $\rho_n$ constitute a set of states in
terms of which we may define the off-diagonal mixed state geometric
phase factors as \cite{filipp03a,filipp03b}
\begin{eqnarray}  
\gamma^{(l)}_{\rho_{j_1}\ldots \rho_{j_l}} & = &  
\Phi \left[ \Tr \left( 
\prod_{a=1}^l U^\parallel(\tau) \sqrt[l]{\rho_{j_a}} \right) \right]   
\label{gammam1}  
\end{eqnarray}  
as long as the trace in the argument of $\Phi$ does not vanish. 
Here, $U^\parallel (t)$ satisfies the parallel transport conditions 
in Eq. (\ref{parallel}) and $l\leq N$. Apparently $\gamma^{(1)}$ 
are the standard mixed state geometric phase factors associated 
with the unitary paths ${\cal M}_n : t \in [0,\tau] \rightarrow 
\rho_n (t)$ in state space. Furthermore, it has been demonstrated 
\cite{filipp03a,filipp03b} a physical scenario for the $l=2$ case 
in terms of two-particle interferometry.
  
As the nonzero eigenvalues of all mixed states under 
consideration are nondegenerate, the parallel transport conditions 
are still given by Eq. (\ref{parallel}) and $U^\parallel (t)$ by  
Eq. (\ref{Uparallel}). Substituting Eq. (\ref{Uparallel}) into  
Eq. (\ref{gammam1}), we obtain the kinematic expression for  
the off-diagonal geometric phase factors for mixed states under  
evolution $U(t)$ as  
\begin{widetext}       
\begin{eqnarray}       
\gamma^{(l)}_{\rho_{j_1} \ldots \rho_{j_l}} & = &  
\Phi\left[\sum_{i_1,\ldots,i_l=1}^N   
\sqrt[l]{\lambda_{i_1-j_1+1}  
\ldots \lambda_{i_N-j_N+1}} \bra{\psi_{i_1}} U(\tau) \ket{\psi_{i_2}}  
\bra{\psi_{i_2}} U(\tau) \ket{\psi_{i_3}} \ldots  
\bra{\psi_{i_N}} U(\tau) \ket{\psi_{i_1}} \right]  
\nonumber\\  
 & & \times \exp \left( -\int_0^\tau \sum_{a=1}^l \bra{\psi_{i_a}}  
U^{\dagger} (t) {\dot U} (t) \ket{\psi_{i_a}} dt \right) ,        
\end{eqnarray}       
\end{widetext}       
where $\lambda_{-p}=\lambda_{N-p}$, $p=0,\ldots,N-1$. 
One may verify that the phase factors 
$\gamma_{\rho_{j_1} \ldots \rho_{j_l}}^{(l)}$ are 
gauge invariant in that they are independent of the choice of 
$U(t)\in {\cal S}$.
  
\section{off-diagonal geometric phases for degenerate density operators}  
\label{sec:degenerate}  
We now generalize the concept of off-diagonal geometric phases of
nondegenerate mixed states to degenerate mixed states. Let 
\begin{eqnarray} 
\rho_1 = \lambda_1 P_{1;1}^{(m_1)} + \ldots + 
\lambda_K P_{1;K}^{(m_K)} ,  
\end{eqnarray}
where the nonzero eigenvalue $\lambda_k$, $k=1,\ldots ,K\leq N$, is 
$m_k$ fold degenerate and $P_{1;k}^{(m_k)}$ the corresponding 
projector of rank $m_k$. The choice $\ket{\psi_1}, \ldots, 
\ket{\psi_N}$ of orthonormal Hilbert space basis and concomitant 
permutation unitarity $W = \ket{\psi_1}\bra{\psi_N} + 
\ket{\psi_N}\bra{\psi_{N-1}} + \ldots + \ket{\psi_2}\bra{\psi_1}$, 
defines a set of noninterfering density operators 
\begin{eqnarray}
\rho_n = \lambda_1 P_{n;1}^{(m_1)} + \ldots + 
\lambda_K P_{n;K}^{(m_K)}, \ n=1,\ldots ,N,   
\end{eqnarray} 
where $P_{n;k}^{(m_k)} =W^{n-1}P_{1;k}^{(m_k)}(W^\dagger)^{n-1}$, 
$n=1,\ldots N$. Note that $P_{n;k}^{(m_k)} \neq P_{n';k}^{(m_k)}$
for $n\neq n'$. 

The parallel transport conditions for a unitary path 
$t \rightarrow \rho_n (t)$ with $\rho_n(0) = \rho_n$, 
could be taken as
\begin{eqnarray}  
P_{n;k}^{(m_k)} U_n^{\parallel\dagger} (t) \dot{U}_n^{\parallel} (t) 
P_{n;k}^{(m_k)} & = & 0 , \ k=1,\ldots K. 
\label{parallel2}       
\end{eqnarray}  
In terms of the orthonormal Hilbert space basis $\ket{\psi_1},\ldots, 
\ket{\psi_N}$, Eq. (\ref{parallel2}) is equivalent to the more 
familiar parallel transport conditions \cite{singh03} $\bra{\psi_{\mu}} 
U_n^{\parallel\dagger} (t) \dot{U}_n^{\parallel} (t) 
\ket{\psi_{\nu}} = 0$, $\forall$ pairs $\ket{\psi_{\mu}},\ket{\psi_{\nu}} 
\in P_{n;k}^{(m_k)}$, and reduces to those of Ref. \cite{sjoqvist00} 
in the nondegenerate case where $K=N$. 

Now, suppose there is a one-parameter family of unitarities  
$\{ U(t) | t\in [0,\tau],U(0)=I \}$ defining the paths  
${\cal M}_n : t\in [0,\tau] \rightarrow \rho_n (t) =  
U(t) \rho_n U^{\dagger}(t)$, $n=1,\ldots,N$. Then, for each $\rho_n$,  
there is an equivalence set ${\cal S}_n$ of unitarities  
$\widetilde{U}_n (t)$ that all realize ${\cal M}_n$, namely  
those of the form  
\begin{eqnarray}       
\widetilde{U}_n (t) =U (t) V_n (t) 
\label{U2a}  
\end{eqnarray}  
with 
\begin{eqnarray}       
V_n (t) = \alpha_{n;1} (t) + \ldots + \alpha_{n;K} (t) ,  
\label{U2b}  
\end{eqnarray}  
where $\alpha_{n;k} (t)$ is unitary on the $k$th degenerate 
subspace, i.e., $\alpha_{n;k} (t)\alpha_{n;k}^{\dagger} (t) = 
\alpha_{n;k}^{\dagger} (t) \alpha_{n;k} (t) = P_{n;k}^{(m_k)}$, 
and $\alpha_{n;k} (0) = P_{n;k}^{(m_k)}$. We can identify 
$U_n^{\parallel} (t) \in {\cal S}_n$  by substituting 
$U_n^{\parallel} (t) = \widetilde{U}_n (t)$ into  Eq. 
(\ref{parallel2}), and obtain 
\begin{widetext} 
\begin{eqnarray}  
\alpha_{n;k}^{\parallel} (t) = P_{n;k}^{(m_k)}  
{\bf T} \exp \left( -\int_{0}^{t} P_{n;k}^{(m_k)} U^{\dagger} (t') 
{\dot U}(t') P_{n;k}^{(m_k)} dt' \right) P_{n;k}^{(m_k)} ,  
\label{U3}    
\end{eqnarray}  
\end{widetext}
where ${\bf T}$ denotes time ordering.  The parallel 
transporting unitary operators for $\rho_n$ may be expressed 
as $U_{n}^\parallel(t) = U(t) V_{n}^\parallel(t)$ with  
supplementary operators $V_{n}^\parallel(t) = 
\alpha_{n;1}^{\parallel} (t) + \ldots + \alpha_{n;K}^{\parallel} (t)$. 

In the presence of degenerate subspaces, some of the parallel 
transporting unitarities $U_n^\parallel(t)$ must be different from 
each other. Thus, unlike the nondegenerate case, where a common 
parallel transporting $U^\parallel(t)$ can always be found, such 
a common operator parallel transporting all the states $\rho_n$ 
does not exist in the degenerate case; we need to take 
$U_1^\parallel(t),\ldots,U_N^\parallel(t)$ to parallel transport 
$\rho_1,\ldots,\rho_N$, respectively. Substituting
$U_{j_a}^\parallel(t)=U(t)V_{j_a}^\parallel(t)$, $a=1,\ldots,l,$ into
Eq. (\ref{gammam1}) for the $a$th $ U^{\parallel}(t)$, we obtain the
off-diagonal geometric phase factors of the degenerate mixed states as
\begin{eqnarray}  
\gamma^{(l)}_{\rho_{j_1} \ldots \rho_{j_l}} & = &  
\Phi \left[ \Tr \left( \prod_{a=1}^l U(\tau) 
V_{j_a}^\parallel (\tau) \sqrt[l]{\rho_{j_a}} 
\right) \right]  
\label{gammam2}  
\end{eqnarray}  
for nonvanishing trace in the argument of $\Phi$.
The phase value calculated in Eq. (\ref{gammam2}) is manifestly gauge
invariant under choice of the set $\{ \widetilde{U}_n (t) \}$, which
shows that it is only dependent upon the paths traced by the
states. We may verify this point by repeating the demonstration of
Ref. \cite{singh03}.
  
Again, once the paths of states are defined by some unitary 
operator $U(t)$, the supplementary operators $V_{j_a}^\parallel(t)$, 
$a=1,\ldots,l$, and therefore $U_{j_a}^\parallel(t)$, may be different 
for each $\rho_{j_a}$. However, if all the states $\rho_{j_a}$ have 
the same degeneracy structure \cite{remark1}, they share a common 
unitary operator $U^\parallel (t)$. For a general set of states 
$\rho_{j_1},\ldots, \rho_{j_l}$, whether a common unitary parallel 
transporting operator exists depends both upon the states and the 
unitarity $U(t)$. In the case where a common parallel transporting 
unitary operator does not exist, two or more unitary operators are 
necessary to calculate the off-diagonal geometric phases. For this 
reason, experimental tests of off-diagonal geometric phases may be 
difficult to implement in such cases. 
 
\section{Example: pseudopure states} 
Let us illustrate the above for pseudopure states, which take 
the form 
\begin{eqnarray} 
\rho_n = \frac{1-\epsilon}{N} I + \epsilon \ket{n} \bra{n} , \ 
n=1,\ldots ,N \geq 2  
\label{nmrstates}   
\end{eqnarray}
with $0< \epsilon \leq 1$, and $\ket{n}$ any $p-$qubit 
state. Such states have a single nondegenerate eigenvector $\ket{n}$ 
with eigenvalue $\big( 1+(N-1)\epsilon \big)/N$ and an $N - 1$ 
fold degenerate eigenvalue $(1-\epsilon)/N$ with eigenprojector 
$I-\ket{n}\bra{n}$. Pseudopure states appear generically 
as input states to standard liquid-state NMR quantum computers, in 
case of which $N=2^p$, $p$ being the number of nuclear spin qubits 
\cite{gershenfeld97}.  

Now, we wish to compute the two lowest order ($l=1$ and $l=2$) 
geometric phases for the pseudopure states $\rho_n$ and $\rho_m$ 
with $\bra{n}m\rangle=0$. Let 
\begin{eqnarray}
U(t)=I-P_{nm}+U_{nm}(t), 
\end{eqnarray}
where $P_{nm}$ is the projection operator onto the subspace spanned 
by $\{ \ket{n},\ket{m} \}$ and $U_{nm}(t)U_{nm}^{\dagger}(t) = 
U_{nm}^{\dagger}(t)U_{nm}(t)=P_{nm}$. In the case of liquid-state 
NMR quantum protocols, one may for example take  
\begin{eqnarray} 
\ket{n} & = & \ket{0}_1 \otimes \ldots \otimes \ket{0}_q \otimes 
\ldots \otimes \ket{0}_p ,  
\nonumber \\    
\ket{m} & = & \ket{0}_1 \otimes \ldots \otimes \ket{1}_q \otimes 
\ldots \otimes \ket{0}_p, 
\end{eqnarray}
where $1\leq q\leq p$
so that $U(t)$ represents single-qubit operations on the $q$th 
nuclear spin. Using Eq. (\ref{U2b}), we obtain the parallel 
transporting unitarities  
\begin{eqnarray}
U_n^{\parallel} & = & U_m^{\parallel} = I - P_{nm} + 
U_{nm} \ket{n}\bra{n} 
\nonumber \\ 
 & & \times e^{-\int_{0}^{t} \bra{n} U_{nm}^{\dagger}(t') 
\dot{U}_{nm} (t') \ket{n} dt'} + U_{nm} \ket{m}\bra{m} 
\nonumber \\ 
 & & \times e^{-\int_{0}^{t} \bra{m} U_{nm}^{\dagger}(t') 
\dot{U}_{nm} (t') \ket{m} dt'} .  
\label{nmrpt}
\end{eqnarray} 
Thus, due to the specific form of $U$, there is a single unitarity
parallel transporting both $\rho_n$ and $\rho_m$. This property 
may be advantageous in experimental tests of the $l=1$ and $l=2$ 
geometric phases for $\rho_n$ and $\rho_m$. Note that a separate 
treatment of the random mixture ($\epsilon = 0$) yields 
$U_n^{\parallel} = U_m^{\parallel} = I$. 

Let us first analyze the $l=1$ geometric phase factors for $\rho_n$ and 
$\rho_m$ in the $\epsilon >0$ case. Inserting Eqs. (\ref{nmrstates}) 
and (\ref{nmrpt}) into Eq. (\ref{gammam2}) yields 
\begin{widetext}
\begin{eqnarray} 
\gamma_{\rho_n}^{(1)} & = & (\gamma_{\rho_m}^{(1)})^{\ast} =  
\Phi \left[ \big( N-2 \big)(1-\epsilon) + 
\eta \big( 1+ (N-1)\epsilon \big) e^{-i\Omega/2} + 
\eta (1-\epsilon) e^{i\Omega/2} \right] 
\nonumber \\ 
 & = & \exp \left( -i \arctan 
\left( \frac{\eta N \epsilon \sin \frac{\Omega}{2}}{(N-2)(1-\epsilon) + 
\eta \big(2+ (N-2)\epsilon \big) \cos \frac{\Omega}{2}} \right) \right) , 
\end{eqnarray}
\end{widetext}
whenever the phase factor is defined. Here, $\Omega$ is the
geodesically closed solid angle on the Bloch sphere defined by
projection of the two-dimensional subspace spanned by the pure state
components $\ket{n}$ and $\ket{m}$ and $\eta = |\bra{n} U_{nm}
\ket{n}| = |\bra{m} U_{nm} \ket{m}|$ is the pure state visibility. In
the absence of noise ($\epsilon = 1$) the $l=1$ geometric phases
become undefined iff $\eta=0$, corresponding to the case where
$\ket{n}$ and $\ket{m}$ are interchanged. When $\eta \neq 0$, we
obtain the standard pure state geometric phases $\gamma_{\rho_n}^{(1)}
= (\gamma_{\rho_m}^{(1)})^{\ast} = e^{-i\Omega/2}$. In the general
case ($0<\epsilon <1$), the nodal points, defined as those points 
in parameter space $(\eta,\epsilon,\Omega)$ where $\gamma_{\rho_n}^{(1)}$ 
and $\gamma_{\rho_m}^{(1)}$ are undefined, are obtained as solutions of  
\begin{eqnarray} 
 & & \left( \big( N-2 \big)(1-\epsilon) + \eta \big(2+ (N-2)\epsilon \big) 
\cos \frac{\Omega}{2} \right)^2 
\nonumber \\ 
 & & + 
\eta^2 N^2 \epsilon^2 \sin^2 \frac{\Omega}{2}  =0 . 
\label{nodal1}  
\end{eqnarray}
For $N=2$ (nondegenerate case), the nodal points are those where 
$\eta =0$, which is consistent with Refs. \cite{filipp03a,filipp03b}, 
and when $\eta \neq 0$ we have 
\begin{eqnarray} 
\gamma_{\rho_n}^{(1)} = (\gamma_{\rho_m}^{(1)})^{\ast} = 
\exp \left( -i\arctan \left( \epsilon 
\tan \frac{\Omega}{2} \right) \right) , 
\end{eqnarray}
which is consistent with Refs. \cite{sjoqvist00,du03} by identifying 
$\epsilon$ as the length of the Bloch vector. For $N\geq 3$ 
a necessary condition for a nodal point is $\Omega/2 = (2k+1)\pi$, 
$k$ integer, and $\eta \neq 0$. For these solid angles, 
Eq. (\ref{nodal1}) reduces to 
\begin{eqnarray} 
\eta = \frac{(N-2)(1-\epsilon)}{2+(N-2)\epsilon} . 
\label{nodal1N>3}
\end{eqnarray}
Physical solutions (i.e., fulfilling $\eta \leq 1$) of this equation 
exist provided 
\begin{eqnarray} 
\epsilon \geq \frac{N-4}{2(N-2)}.
\label{epi} 
\end{eqnarray}
Thus, for $N\geq 5$, there is a lower bound on the amount of noise 
for the existence of nodal points, i.e., in too noisy input states 
the $l=1$ geometric phases are always well-defined for this $U$. 

Let us now turn to the $l=2$ geometric phase for $\rho_n$ and 
$\rho_m$. First notice that $\gamma_{\rho_n \rho_m}^{(2)} = 
\gamma_{\rho_m \rho_n}^{(2)}$, which is due to the symmetry of 
trace under cyclic permutations. Explicit calculation yields  
\begin{eqnarray} 
\gamma_{\rho_n \rho_m}^{(2)} & = & 
\Phi \Big[ (N-2) (1-\epsilon) +  
\big( 2 + (N-2)\epsilon \big) (-1+\eta^2)  
\nonumber \\ 
 & & + 2\eta^2 \sqrt{ (1-\epsilon)( 1 + (N-1) \epsilon )} \ 
\cos \Omega \Big] .  
\label{2nmr}
\end{eqnarray}
Thus, $\gamma_{\rho_n \rho_m}^{(2)} = \pm 1$, changing sign across 
a nodal surface in the parameter space $(\epsilon,\eta,\Omega)$. 
This nodal surface is defined by vanishing argument of $\Phi$ in 
Eq. (\ref{2nmr}), i.e., the solution of 
\begin{eqnarray}
\eta^2 & = & \frac{2(N-2)\epsilon - N+4}{2+(N-2)\epsilon + 
2\sqrt{(1-\epsilon)\big( 1 + (N-1)\epsilon \big)} \ \cos \Omega} . 
\nonumber \\ 
\label{2nodalnmr}
\end{eqnarray}  
If $\cos \Omega \geq 0$, solutions fulfilling $\eta \leq 1$ always 
exist for all $N$ as long as Eq. (\ref{epi}) is satisfied. In the 
case where $\cos \Omega <0$, however, $0 \leq \eta^2 \leq 1$ yields
\begin{eqnarray} 
\frac{N-4}{2(N-2)} \leq \epsilon \leq 
\frac{(N-2)^2 - 4\cos^2 \Omega}{4(N-1)\cos^2 \Omega + (N-2)^2}.
\end{eqnarray}
Thus, there are nodal points only for the amount of noise satisfying
the above equation. In the limiting case $N=2$, the upper bound on
$\epsilon$ is negative and thus no nodal points exist, which is 
consistent with Refs. \cite{filipp03a,filipp03b}.

\begin{figure}[htbp]
\centering
\includegraphics[width=80mm]{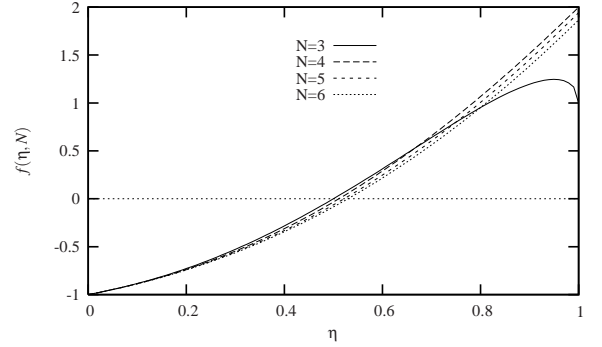}
\caption{Common nodal points of $\gamma_{\rho_n \rho_m}^{(2)}$ and
$\gamma_{\rho_n}^{(1)}=(\gamma_{\rho_m}^{(1)})^{\ast}$ for pseudopure 
states $\rho_n$ and $\rho_m$ are given by vanishing $f(\eta,N)$, which 
is depicted for $N=3,4,5,6$. Thus, except for the qubit case ($N=2$), 
there may arise situations where the $l=1$ and $l=2$ geometric phases 
are undefined simultaneously.}
\label{f1}
\end{figure}

Now we wish to check $\gamma_{\rho_n \rho_m}^{(2)}$ at the nodal 
points of $\gamma_{\rho_n}^{(1)}=(\gamma_{\rho_m}^{(1)})^{\ast}$. 
For $N=2$ we obtain that the nodal points for $l=1$ and $l=2$ 
never coincide. For $N\geq 3$, insert $\Omega/2=(2k+1)\pi$, 
$k$ integer, into Eq. (\ref{2nodalnmr}) and eliminate $\epsilon$ 
by using Eq. (\ref{nodal1N>3}), yielding  
\begin{eqnarray} 
\eta^2 + \eta -1 + \frac{2\eta^2}{N-2} \sqrt{\eta (N-2-\eta)} \equiv 
f(\eta,N) = 0, 
\end{eqnarray}
the solutions of which are the common nodal points of 
$\gamma_{\rho_n \rho_m}^{(2)}$ and $\gamma_{\rho_n}^{(1)}$. In
Fig. \ref{f1} we have depicted $f(\eta,N)$ for $N=3,4,5,6$ and we
conclude that there are coinciding nodal points for $N > 2$. For 
these rare parameter values one has to resort to $l\geq 3$ geometric 
phases to retain some information about the geometry of the path in 
state space. 

To measure $\gamma_{\rho_1 \rho_2}^{(2)}$ we add an ancillary system
$a$ and consider pure states $\ket{\Psi_n} \in \mathcal{H}_s
\otimes \mathcal{H}_a$, $\dim \mathcal{H}_a = N$, such that 
the partial trace $\Tr_a \ket{\Psi_n} \bra{\Psi_n} = \rho_n$.
For pseudopure states, the purifications take the form 
\begin{eqnarray}
\ket{\Psi_n} & = & \sqrt{\frac{1-\epsilon}{N}} \sum_{k \neq n} 
\ket{k} \otimes \ket{\varphi_k} 
\nonumber \\ 
 & & + \sqrt{\epsilon + \frac{1-\epsilon}{N}} \ket{n} \otimes 
\ket{\varphi_n} , 
\end{eqnarray}
where $\ket{\varphi_1},\ldots ,\ket{\varphi_N}$ is an orthonormal
basis of $\mathcal{H}_a$. $\gamma_{\rho_1 \rho_2}^{(2)}$ may be
measured interferometrically as a shift in the interference pattern
\begin{eqnarray}
\mathcal{I} \propto \left| U_s \otimes U_a \ket{\Psi_n} + 
V_s \otimes V_a \ket{\Psi_n} \right|^2
\label{intensity}
\end{eqnarray}
by choosing $U_s = e^{i\chi} W^{m-n}$, $V_s = U_n^{\parallel}$, 
$U_a = W^{m-n}$, and $V_a = (U_n^{\parallel})^{\textrm{T}}$, 
T being transpose with respect to the ancilla basis 
$\ket{\varphi_1},\ldots ,\ket{\varphi_N}$ and $\chi$ a variable U(1) 
phase shift \cite{filipp03b}.   

In liquid-state NMR, such a test could in principle be 
implemented as follows. By adding an `auxiliary' nuclear spin qubit 
pair in the state $\ket{0} \otimes \ket{0}$, prepare the total input   
\begin{eqnarray} 
\ket{\Gamma} = \ket{0} \otimes \ket{0} \otimes \ket{\Psi_n} .  
\end{eqnarray}  
Apply ${\bf H} \otimes {\bf H} \otimes I_s \otimes I_a$, 
${\bf H} \otimes {\bf H}$ being Hadamard on each auxiliary qubit, 
followed by the conditional transformation 
\begin{eqnarray}
U_c & = & 
\ket{0} \bra{0} \otimes \ket{0} \bra{0} \otimes U_s \otimes U_a 
\nonumber \\ 
 & & + \ket{1} \bra{1} \otimes \ket{1} \bra{1} \otimes V_s \otimes V_a  
\end{eqnarray}  
with $U_s,U_a,V_s,$ and $V_a$ chosen as above. Finally, apply 
another Hadamard transform on the auxiliary qubits and measure 
the output in the $00$ channel. The resulting intensity should 
confirm Eq. (\ref{intensity}). 

\section{Conclusions} 
We have put forward a kinematic approach to the off-diagonal geometric
phases of pure states \cite{manini00} and nondegenerate mixed states
\cite{filipp03a,filipp03b}. The kinematic approach to off-diagonal 
geometric phases presented in this paper emphasizes the path in state
space as the primary concept for the geometric phase and provides a
tool to efficiently calculate off-diagonal geometric phases for
arbitrary unitary operators and arbitrary density matrices. We have
extended the concept of off-diagonal geometric phases of nondegenerate
mixed states to degenerate mixed states. This kind of extention is
essential since the geometric phase of degenerate mixed
states \cite{singh03} is undetermined in the case where the
interference visibility between the initial and final states vanishes
leading to an interesting nodal point structure in the experimental
parameter space that could be monitored in a history-dependent
manner. On the other hand, the notion of off-diagonal geometric phase
put forward for nondegenerate mixed state in Refs. 
\cite{filipp03a,filipp03b} breaks down in the case of degenerate 
mixed states, while the latter case does really include the most
general quantum states and therefore this extention is physically
important. Our extension makes it possible to obtain information of
the geometry along the path in state space when the standard notion of
geometric phase for degenerate density operators is undetermined.
We have argued that liquid-state NMR protocols constitute a possible 
physical scenario for test of off-diagonal geometric phases for 
degenerate mixed quantal states. 

\vskip 0.3 cm     
The work by Tong was supported by NUS Research Grant No.   
R-144-000-071-305. E.S. acknowledges partial financial support  
from the Swedish Research Council. S.F. acknowledges support from  
the Austrian Science Foundation, Project No. F1513.


\begin{thebibliography}{99}
\bibitem{remark1}  
By saying that the unitarily connected mixed states $\rho_1,\ldots,
\rho_l$ have the same degeneracy structure, we mean that an 
eigenvector $\ket{\psi}$ of $\rho_n$ corresponding to an $m_i-$fold 
degenerate eigenvalue $\lambda_{i}$ is also an eigenvector of 
$\rho_{n'\neq n}$ corresponding to an $m_i-$fold degenerate eigenvalue
$\lambda_{j}$. Note that $\lambda_{i} \neq \lambda_{j}$, 
in general.
\end{thebibliography}
\end{document}